\documentclass[aps,prl,twocolumn,groupedaddress,amsmath,amssymb]{revtex4-1}
\usepackage{graphicx}  
\usepackage{dcolumn}   
\usepackage{bm}        
\usepackage{verbatim}   
\usepackage{amssymb}
\usepackage{amsfonts}
\usepackage{amsmath}
\usepackage{graphicx}
\usepackage{bm}
\usepackage{subfig,tikz}
\usepackage{wrapfig}
\usepackage{xspace}
\usepackage{enumerate}
\newtheorem{theorem}{Theorem}

\newtheorem{corollary}[theorem]{Corollary}

\newcommand{\R}{\mathbb R}

    \usepackage{xargs} 
    \newcommand{\iso}{\cong}
    
    \newcommand{\D}{\mathcal{D}}
    
    \newcommand{\sub}{\subset}

    \newcommand{\zb}{\mathbf{z}}
    \newcommand{\yb}{\mathbf{y}}

    \DeclareMathOperator{\spn}{span}

    \renewcommand{\v}{\mathbf{v}}

    \renewcommand{\j}{\mathbf{j}}
    \newcommand{\e}{\mathbf{e}}
    \newcommand{\xb}{\mathbf{x}}

    \newcommand{\w}{\mathbf{w}}

    \newcommand{\Z}{\mathbb{Z}}

\begin{document}
\title{Black Lenses in Kaluza-Klein Matter}

\author{Marcus A. Khuri}
\email{khuri@math.sunysb.edu}
\author{Jordan F. Rainone}
\email{jordan.rainone@stonybrook.edu, rainonej@gmail.com}
\address{Department of Mathematics, Stony Brook University, Stony Brook, NY 11794, USA}


\begin{abstract}  \noindent
We present the first examples of formally asymptotically flat black hole solutions with horizons of general lens space topology $L(p,q)$.
These 5-dimensional static/stationary spacetimes are regular on and outside the event horizon for any choice of relatively prime integers $1\leq q<p$, in particular conical singularities are absent. They are supported by Kaluza-Klein matter fields arising from higher dimensional vacuum solutions through reduction on tori. The technique is sufficiently robust that it leads to the explicit construction of regular solutions, in any dimension, realising the full range of possible topologies for the horizon as well as the domain of outer communication, that are allowable with multi-axisymmetry. Lastly, as a by product, we obtain new examples of regular gravitational instantons in higher dimensions. 
\end{abstract}
\maketitle


What are the possible shapes of a black hole? Fifty years ago, Hawking \cite{Hawking} provided an answer to this fundamental question in spacetime dimension 4, with his horizon topology theorem. This result asserts that cross-sections
of the event horizon, in asymptotically flat stationary black hole spacetimes satisfying the dominant energy condition, must be topologically a 2-sphere $S^2$. In 2002, Emparan-Reall \cite{EmparanReall} discovered the first regular asymptotically flat non-spherical black hole, a 5-dimensional black ring with horizon topology $S^1 \times S^2$. Not only did this give impetus to the question above, but it also showed that the traditional black hole no hair theorem is false in higher dimensions \cite{ER1}. Shortly thereafter, Galloway-Schoen \cite{GS} generalized Hawking's theorem to higher dimensions, stating that horizon cross-sections must be of positive Yamabe invariant. This condition is equivalent to the underlying topology admitting a metric of positive scalar curvature, and leads to a concise list of possible horizon topologies in five dimensions \cite{Ga}. Namely, orientable horizons in this setting must be either a quotient of the 3-sphere $S^3$ (spherical space form), the ring $S^1 \times S^2$, or a finite connected sum thereof. Further restrictions are possible in the case of extreme black holes \cite{KhuriWoolgarWylie}, in particular all but one connected sum can be ruled out.

The basic question of whether each topology on the list is achieved by a black hole solution has remained unresolved.
The totality of non-spherical black holes found to date consists of the ring $S^1 \times S^2$ \cite{EmparanReall,PS}, and the lens spaces $L(p,1)$ discovered initially by Kunduri-Lucietti when $p=2$ \cite{KunduriLucietti} and extended to all positive integers $p$ by Tomizawa-Nozawa \cite{TN}, see also \cite{KL}. While the ring is a vacuum solution, the lenses are solutions of minimal supergravity \cite{Cremmer}. Moreover, there is evidence that suggests regular asymptotically flat vacuum black lenses do not exist \cite{LT}, and proposals to balance black lenses in a bubble of nothing \cite{Witten} have been unsuccessful \cite{TS}
(although it is possible for black rings \cite{AEV}).

Symmetry yields further restrictions on topology. Indeed, the rigidity theorem \cite{HIW,IM} guarantees that generically stationary black holes come with at least one rotational symmetry, and in fact almost all known solutions in 5-dimensions have $U(1)^2$ rotational symmetry; see \cite{KL} for recent examples admitting only $U(1)$ symmetry. In this setting of bi-axisymmetry, the list of possible horizon topologies reduces to the sphere $S^3$, the ring $S^1 \times S^2$, and the lens spaces $L(p,q)\cong S^3 /\mathbb{Z}_p$ for any choice of relatively prime integers $1\leq q<p$. It is also possible to classify the list of possible domain of outer communication (DOC) topologies \cite{HHI,KKRW,KMWY} in this regime. Namely, the compactified Cauchy surfaces within the DOC must either be the 4-sphere $S^4$, a connected sum of $S^2 \times S^2$'s, or in the non-spin case a connected sum of complex projective planes $\mathbb{CP}^2$ and $\overline{\mathbb{CP}}^2$. There have been a number of attempts to realize all the topologies in these lists, however they have all suffered from the presence of either naked singularities \cite{CT,E}, conical singularities on the axes \cite{KWY,KKRW,LT}, or closed timelike curves \cite{TM}, when trying to implement the more complicated configurations. In spacetime dimensions greater than 5 very little is known, although a systematic study of static vacuum generalized Weyl solutions was given in \cite{ERstatic} and stationary vacuum solutions with possible conical singularities were produced in \cite{KKRW}.

The purpose of this note is the following. We show that all possible horizon topologies, and DOC topologies from the classification, including all combinations of multiple horizon configurations, are realized by regular formally asymptotically flat black hole solutions. In particular, this includes the first examples of general lens space topology $L(p,q)$, involving a topology change between the horizon and asymptotic end.
These black holes can be either static or stationary, and are supported by Kaluza-Klein matter fields in that they arise from higher dimensional vacuum solutions through reduction on tori. The methods also extend to all higher dimensions, allowing for the construction of solutions, realising the full range of possible topologies for the horizon as well as the DOC, that are compatible with multi-axisymmetry in which the orbit space is 2-dimensional. Furthermore, as a by product, we obtain new examples of regular gravitational instantons \cite{Hawking0} in higher dimensions. 

The basic strategy consists of the following steps. Given the desired DOC $\mathcal{M}^{n+3}$ for a $(n+3)$-dimensional static/stationary spacetime admitting $U(1)^n$ symmetry with $n\geq 1$, we show how to encode its topology in a higher dimensional DOC $\tilde{\mathcal{M}}^{n+3+k}$ having a relatively simple topological structure. On this higher dimensional spacetime manifold, we solve the static/stationary vacuum Einstein equations with $U(1)^{n+k}$ symmetry, and take advantage of the simple topology to balance any conical singularities (choose parameters to achieve a cone angle of zero). A dimensional reduction, or quotient procedure, is then carried out in order to obtain a regular solution with Kaluza-Klein matter on the original topology $\mathcal{M}^{n+3}$.

Due to global hyperbolicity, the topology of the spacetimes considered here will always be of the form $\mathcal{M}^{n+3}=\R\times M^{n+2}$. The time slice $M^{n+2}$ is assumed to admit an effective action by the torus $T^n=U(1)^n$, and hence the quotient map $M^{n+2}\to M^{n+2}/T^n$ exhibits $M^{n+2}$ as a $T^n$-bundle over a $2$-dimensional base space with any degenerate fibers occurring on the boundary. In particular, while
fibers over interior points are $n$-dimensional, fibers over boundary points can be $(n-1)$ or $(n-2)$-dimensional. Those points where the fiber is $(n-1)$-dimensional are called \emph{axis rods}, and the points with an $(n-2)$-dimensional fiber are discrete and called \emph{corners}. Consistency with topological censorship demands that the base space $M^{n+2}/T^n$ is homeomorphic to a half plane $\R_{+}^2$~\cite{HY}. 

The entire topology of $M^{n+2}$ may be recorded in the boundary $\partial\R^2_+$ of this half-plane. This is achieved by dividing it into disjoint intervals separated by corners or \emph{horizon rods} (assumed to be finite) where the fibers do not degenerate. 
Associated to each axis rod interval $\Gamma_i\sub \partial \R^2_+$ is a vector $\v_i\in \Z^n$ referred to as the \emph{rod structure}, which determines the 1-dimensional isotropy subgroup $ \R \cdot \v_i + \Z^n \sub \R^n / \Z^n \iso T^n$ 
for the action of $T^n$ on points that lie over $\Gamma_i$. See \cite{ERstatic,Harmark,KKRW} for further discussion concerning rod structures. We then have
\begin{equation}\label{aouhf}
M^{n+2}\iso (\R^2_+\times T^n)/\sim,
\end{equation} 
where the equivalence relation $\sim$ is given by $(\mathbf{p},\pmb\phi) \sim (\mathbf{p}, \pmb\phi + \lambda \v_i )$ with $\mathbf{p}\in \Gamma_i$, $\lambda\in \R/\Z$, and $\pmb\phi \in T^n$. 
Together, the rod structures form a ($n$-dimensional) \emph{rod data set} $\D = \{(\v_1, \Gamma_1), \dots, (\v_k, \Gamma_k)\}$ which enshrines the topology of the DOC. Rod data sets may be chosen arbitrarily except for an \emph{admissibility condition} when $n\geq 2$ that guarantees the total space is a manifold \cite[Section 3]{KKRW}, namely if rod structures $\v_i, \v_{i+1}$ arise from neighboring rods separated by a corner then the 2nd determinant divisor
\begin{equation*}
\operatorname{det}_2(\v_i,\v_{i+1})=\mathrm{gcd}\{|Q_{\j_1 \j_2}|\}
\end{equation*}
is 1, 
where $Q_{\j_1 \j_2}$ is the determinant of the $2\times 2$ minor obtained from the matrix defined by the column vectors $\v_{i},\v_{i+1}$ by picking rows $\j_1$ and $\j_2$. We may assume without loss of generality that each rod structure $\v_i$ is primitive, in the sense that its components are relatively prime $\mathrm{gcd}(v_i^1,\ldots,v_i^n)=1$, since this does not change the associated isotropy subgroup.

Given a topology $\mathcal{M}^{n+3}$ that we wish to realize as the DOC for a static/stationary solution of the Einstein equations, and which is characterized by a $n$-dimensional admissible rod data set $\mathcal{D}$, the first goal is to encode this into a higher dimensional rod data set having a simpler structure. To this end, we define a new $(n+k)$-dimensional rod data set $\tilde{\mathcal{D}}=\{(\tilde{\v}_1, \Gamma_1), \dots, (\tilde{\v}_k, \Gamma_k)\}$ by $\tilde{\v}_i=\bar{\v}_i +\e_{n+i}$ for $i=1,\ldots,k$, where $\e_j$ is an element of the standard basis for $\Z^{n+k}$ having 1 in position $j$ and zeros elsewhere, and $\bar{\v}_i=(\v_i,\mathbf{0})\in \Z^{n+k}$. Note that each vector $\tilde{\v}_i$ is primitive since the same is true for $\v_i$, and similarly since $\det_2(\tilde{\v}_i, \tilde{\v}_j)$ divides $\det_2(\v_i, \v_j)$ the data set $\tilde{\mathcal{D}}$ inherits the admissibility property from $\mathcal{D}$. In particular, the analogous quotient $\tilde{M}^{n+k+2}$ as in \eqref{aouhf} defined with respect to $\tilde{\mathcal{D}}$ yields an $(n+k+2)$-dimensional manifold admitting an effective action by $U(1)^{n+k}$, which will serve as a DOC time slice for a static/stationary spacetime.

We claim that topologically the new higher dimensional manifold is relatively simple, in that it is the product of a torus with a connected sum consisting of products of spheres, and can be described by a rod structure having only standard basis elements. To see this, we note that changing coordinates on the torus fibers $T^{n+k}\iso \R^{n+k}/\Z^{n+k}$ does not change the topology of $\tilde{M}^{n+k+2}$. Such a coordinate change may be described by a matrix $U\in SL(n+k, \Z)$
defined by $U(\e_j)=\e_j$ for $j=1, \dots, n$ and $U(\tilde{\v}_i) = \e_{n+i}$ for $i=1, \dots, k$ so that
\begin{equation*}
    U^{-1} = \begin{bmatrix}
        I_n & V \\
        0 & I_k
    \end{bmatrix},
\end{equation*}
where $I_n$ is the $n\times n$ identity matrix, $I_k$ is the $k\times k$ identity matrix, and $V$ is the $n\times k$ matrix consisting of the rod structures $[\v_1, \dots, \v_k]$.  Thus, after this coordinate change the rod data set becomes $\tilde{\mathcal{D}}'=\{(\e_{n+1}, \Gamma_1), \dots, (\e_{n+k}, \Gamma_k)\}$, so that according to \cite[Theorem 3.4]{MG} (see also the discussion in the proof of \cite[Theorem 2 (iv)]{KRWY}) the topology of the compactified manifold $\tilde{M}_c^{n+k+2}$ is given by
\begin{equation}\label{qaohfoiqhoih}
\left[\overset{k-3}{\underset{\ell=1}{\#}}\ell {k-2 \choose \ell+1}  S^{2+\ell}\times S^{k-\ell}\right]\times T^{n}
\end{equation}
for $k\geq 4$, whereas for $k=2, 3$ the topology is $S^4 \times T^{n}$, $S^5 \times T^{n}$ respectively. In this expression binomial coefficients are used to indicate the number of times the connected sum is taken for each $\ell$. Here the compactified manifold is obtained from $\tilde{M}^{n+k+2}$ by filling-in each horizon as well as infinity with a 4-ball cross a torus $B^4 \times T^{n+k-2}$ (attach along common boundaries), since horizons are characterized by neighboring rod structures $\e_i,\e_{i+1}$ showing that they have topology $S^3 \times T^{n+k-2}$, and similarly for the cross-section at infinity.

We will now solve the Einstein equations on $\tilde{\mathcal{M}}^{n+k+3}=\mathbb{R}\times \tilde{M}^{n+k+2}$ to obtain a regular static vacuum spacetime realizing this DOC topology; at the end it will be explained how to similarly obtain the rotating stationary analogues. An ansatz, studied initially by Emparan-Reall \cite{ERstatic}, will be imposed that restricts the metric along the torus fibers to be given as a diagonal matrix function yielding the following form for the spacetime metric
\begin{equation*}\label{aopfijjhh9owq}
\tilde{\mathbf{g}}=-\rho^2 e^{-\sum_{i=1}^{n+k}u_i}dt^2+e^{2\alpha}(d\rho^2+dz^2)+\sum_{i=1}^{n+k}e^{u_i} \left(d\psi^i \right)^2,
\end{equation*}
where all coefficients depend only on $\rho>0$, $z$ which parameterize the orbit space half-plane $\mathbb{R}_+^2$, and the Killing fields $\partial_{\psi^i}$ generate the $U(1)^{n+k}$ rotational isometries with $0\leq \psi^i<2\pi$. In this setting the static vacuum Einstein equations reduce to finding $n+k$ axisymmetric harmonic functions $u_i$ on $\mathbb{R}^3 \setminus\Gamma$, where $\mathbb{R}^3$ is parameterized in cylindrical coordinates $(\rho,z,\phi)$ with $0\leq \phi<2\pi$ and $\Gamma$ denotes the $z$-axis. The remaining metric coefficient $\alpha$ may be solved by quadrature, using harmonicity of the $u_i$ as an integrability condition 
\begin{align*}\label{aqonihoinhq}
\begin{split}
\alpha_\rho =&
\tfrac\rho8 \left[ (\Sigma u_{i,\rho})^2 - (\Sigma u_{i,z})^2 + \Sigma \left(u_{i,\rho}^2 -u_{i,z}^2 \right)-\tfrac{4}{\rho}\Sigma u_{i,\rho} \right], \\
\alpha_z =& \tfrac\rho4 \left[ (\Sigma u_{i,\rho})(\Sigma u_{i,z}) + \Sigma u_{i,\rho}u_{i,z} -\tfrac2\rho \Sigma u_{i,z}\right].
\end{split}
\end{align*}
Observe that with the spacetime metric ansatz, axes can only exhibit rod structures of type $\mathbf{e}_i$, $i=1,\ldots,n+k$. Moreover, for an axis rod $\Gamma_l$ having the rod structure $\mathbf{e}_l$, we find that the corresponding logarithmic angle defect \cite{KRWY} is given by
\begin{equation}\label{angledefect}
\mathbf{b}_l =  \lim_{\rho \rightarrow 0}  \left(\log \rho +\alpha-\frac{1}{2}u_l\right).
\end{equation}
Recall that nonzero logarithmic angle defect is associated with a force exerted by the axis, which arises from the geometric singularity \cite[Section 5]{W}. It is known that the limit \eqref{angledefect} is constant along the axis $\Gamma_l$, a fact which may be derived from the equations defining $\alpha$ and the asymptotic expansion of $u_l$ with respect to $\rho$. In a more general setting this was established in \cite[Section 3.1]{Harmark}.

The harmonic functions will be taken as potentials for a uniform charge distribution along associated axis rods; note, however, that $\tilde{\mathbf{g}}$ will solve the vacuum equations so Maxwell fields are not present. More precisely, suppose that the axis rods consist of the intervals $\Gamma_1 =(-\infty, b_1]$, $\Gamma_i =[a_i,b_i]$ for $i=2,\ldots,k-1$, and $\Gamma_{k}=[a_k,\infty)$, where $a_i<b_i\leq a_{i+i}<b_{i+1}$ for each $i$.
We then set 
\begin{equation*}\label{qoihroiqhh}
u_{n+i}=\log(r_{a_i}-z_{a_i})-\log(r_{b_i}-z_{b_i})
\end{equation*}
for $i=2,\ldots,k-1$, and 
\begin{equation*}
u_{n+1}=2\log\rho-\log(r_{b_1}-z_{b_1}), \text{ }
u_{n+k}=\log(r_{a_k}-z_{a_k}), 
\end{equation*}
where $r_a = \sqrt{\rho^2+(z-a)^2}$ and $z_a = z-a$. Each of the individual logarithm expressions is harmonic. Furthermore, observe that the functions with $i=2,\ldots,k-1$ are negatively valued and satisfy the following properties: $u_{n+i} \thicksim 2\log\rho$ near $\Gamma_i$, and
$u_{n+i}=(a_i-b_i)/r + O(r^{-2})$ as $r\to\infty$. The remaining functions are set to $u_j=0$, $j=1,\ldots, n$ since they are not linked to axis rods. Clearly then, these harmonic functions guarantee that the desired rod data set $\tilde{\mathcal{D}}'$ is achieved through the metric $\tilde{\mathbf{g}}$.

The spacetime $(\tilde{\mathcal{M}}^{n+k+3},\tilde{\mathbf{g}})$ has the desired topology, satisfies the static vacuum equations, and is asymptotically Kaluza-Klein in the sense that when distances are very far from the horizon the spacetime is approximately the product of 5-dimensional Minkowski space with a flat torus of dimension $n+k-2$. However, it may possess conical singularities along axis rods.
Nevertheless, due to the diagonal matrix structure of the torus fiber metrics, any conical singularity along an axis rod $\Gamma_i$ may be resolved by adding an appropriate constant to the associated harmonic function $u_{n+i}\mapsto u_{n+i}+c_i$, where the constant $c_i$ is chosen to ensure that the logarithmic angle defect $\mathbf{b}_i=0$ in \eqref{angledefect}. This translation in the harmonic functions does not alter any of the properties listed above for the spacetime. We note that a related balancing procedure was employed by 
Emparan-Reall in \cite{ERstatic} for certain examples; it was also used more recently in \cite{KRWY,KWY1}. Furthermore, absence of conical singularities leads to full regularity of the spacetime metric, a fact which may be established analogously to \cite[Section 5.1]{KWY1}. 
A similar procedure may be used to produce regular stationary vacuum solutions having the same rod data set, with prescribed angular momenta for each black hole, by utilizing the results of \cite{KKRW}; although we do not pursue this here.

We now record two auxiliarly results, concerning the ability to achieve certain DOC topologies, that are consequences of the above arguments. Notice that the assumption $n\geq 1$ is not required when constructing the higher dimensional static vacuum spacetime, and this leads to the following statement.

\begin{theorem}
For each pair of integers $n\geq 0$ and $k\geq 2$, the compactified domain of outer communication topology $\tilde{M}^{n+k+2}_c$ given by \eqref{qaohfoiqhoih}, is realized by time slices of a regular, asymptotically Kaluza-Klein (or asymptotically flat when $n=0$, $k=2$), static vacuum solution with up to $k-1$ horizons of topology $S^{3}\times T^{n+k-2}$.
\end{theorem}

In fact, the construction proceeds just as well if no horizons are present. In this case, the $z$-axis consists entirely of axis rods. Furthermore, in this case, for some constant $c$ the function $\sum_{i=1}^{n+k}u_i-2\log\rho-c$ is harmonic on $\mathbb{R}^3\setminus\Gamma$, tends to zero at infinity, and remains bounded upon approach to $\Gamma$. Therefore, a version of the maximum principle (or Weinstein Lemma \cite[Lemma 8]{Weinstein}) shows that this function vanishes identically, that is $\sum_{i=1}^{n+k}u_i=2\log\rho +c$ and hence the static potential is constant. It follows that the time slice is a complete Ricci flat Riemannian manifold, yielding new examples of higher dimensional gravitational instantons.

\begin{corollary}
For each pair of integers $n\geq 0$ and $k\geq 2$, the topology $\tilde{M}^{n+k+2}_c$ gives rise to a gravitational instanton. More precisely, on the complement of a $B^4 \times T^{n+k-2}$ this manifold admits a regular, complete, Ricci flat Riemannian metric which is asymptotically Kaluza-Klein (or asymptotically flat when $n=0$, $k=2$).
\end{corollary}

In order to proceed with the original problem of realizing a static solution on the given topology $\mathcal{M}^{n+3}$ with rod data set $\mathcal{D}$, we will perform a dimensional reduction (or quotienting procedure) on the constructed static spacetime $\tilde{\mathcal{M}}^{n+k+3}$ having rod data set $\tilde{\mathcal{D}}$. First note that the static vacuum metric $\tilde{\mathbf{g}}$ is expressed above with coordinates $\psi^i$, on the torus fibers, that yield the simplistic rod data set $\tilde{\mathcal{D}}'$ in terms of standard basis vectors, however we may change back to the original coordinates $\phi^i$ in which the rod data set is given by $\tilde{\mathcal{D}}$. This is achieved with the unimodular matrix $U=(U^i_j)$ through the relation $\psi^i=U^i_j \phi^j$. It follows that in these coordinates
\begin{equation*}\label{aopfijjhh9owq11}
\tilde{\mathbf{g}}=-\tilde{f}^{-1}\rho^2 dt^2+\tilde{f}^{-1}e^{2\sigma}(d\rho^2+dz^2)+\sum_{i,j=1}^{n+k}\tilde{f}_{ij}d\phi^i d\phi^j,
\end{equation*}
where $(\tilde{f}_{ij})=U^T \mathrm{diag}(e^{u_1},\ldots,e^{u_{n+k}})U$, $\tilde{f}=\det (\tilde{f}_{ij})$, and $2\sigma=2\alpha+\log\tilde{f}$ where $\alpha$ is defined above \eqref{angledefect}.

The reduction procedure will be carried out using a $k$-dimensional torus whose action is free (devoid of fixed points) on $\tilde{\mathcal{M}}^{n+k+3}$. In fact, the desired subtorus action is defined by 
\begin{equation*}
T^k \iso\spn_\R\{\e_{n+1}, \dots, \e_{n+k}\}/\spn_\Z\{\e_{n+1}, \dots, \e_{n+k}\}. 
\end{equation*}
To confirm that this is indeed free, we will show that the circle action of $\R/\Z \cdot \w\sub \R^{n+k}/\Z^{n+k}$ is free for any primitive vector $\w \in \spn_\Z\{\e_{n+1}, \dots, \e_{n+k}\}$. Proceeding by contradiction, assume that for some $\w$ the action is not free.  Since fixed points can only occur at axis rods or corners, this implies that for some
$i\in \{1, \dots, k-1\}$ there are $\lambda, \alpha, \beta\in\mathbb{R}$ with $0<\lambda\leq 1$, and $\zb\in\mathbb{Z}^{n+k}$, such that $\lambda \w+\zb =\alpha \tilde{\v}_i +\beta\tilde{\v}_{i+1}$. 
If $\lambda$ is irrational then utilize the transformation matrix $U$ to obtain the equation $\lambda U\w +U\zb =\alpha \e_{n+i} +\beta \e_{n+i+1}$, and observe that then all components of $U\w$ and $U\mathbf{z}$ vanish except possibly those in the $n+i$, $n+i+1$ positions. Writing $U\mathbf{z}$ as a linear combination of $\e_{n+i}$, $\e_{n+i+1}$, and applying the inverse transformation then shows that $\w=\alpha ' \tilde{\v}_i +\beta' \tilde{\v}_{i+1}$. However, this is impossible since $w^j=0$ for $j=1,\ldots, n$ while $\v_i$ and $\v_{i+1}$ are linearly independent. It follows that $\lambda$ is rational, and hence so are $\alpha$ and $\beta$.

We may now find relatively prime integers $\mathrm{a},\mathrm{b},\mathrm{c},$ and $1<\mathrm{d}\leq \mathrm{m}$, such that $\lambda=\mathrm{d}/\mathrm{m}$ and 
\begin{equation*}\label{qohroihqoh113}
\mathrm{c} \tfrac{\mathrm{d}}{\mathrm{m}} \w + \mathrm{c} \zb = \mathrm{a} \tilde{\v}_i + \mathrm{b} \tilde{\v}_{i+1}.
\end{equation*}
Let $\xb\in \Z^n$ and $\yb, \mathbf{w}_k  \in \Z^{k}$ be such that $\zb = (\xb,\yb)$ and 
$\mathbf{w}=(\mathbf{0},\mathbf{w}_k)$, then this equation splits into two parts
\begin{equation*}
    \mathrm{c} \xb = \mathrm{a} \v_i + \mathrm{b} \v_{i+1},\quad\text{ }
    \mathrm{c}\mathrm{d}\w_k =\mathrm{m}(\mathrm{a} \e_i + \mathrm{b} \e_{i+1}- \mathrm{c} \yb).
\end{equation*}
Clearly $\mathrm{m}$ cannot divide $\mathrm{d}$, and also $\mathrm{m}$ cannot divide every component of $\w_k=(w^{n+1}, \dots, w^{n+k})$ since $\w$ is primitive. It follows that $\mathrm{m}$ must divide $\mathrm{c}$, and thus $\mathrm{c}=\mathrm{m}\mathrm{c}'$ for some integer $\mathrm{c}'$. Since the rod structures making up $\mathcal{D}$ satisfy the admissibility condition, we then have 
\begin{equation*}
    \mathrm{b} = \operatorname{det}_2(\v_i, \mathrm{a}\v_i + \mathrm{b}\v_{i+1})
    = \mathrm{m}\operatorname{det}_2(\v_i, \mathrm{c}'\xb).
\end{equation*}
Hence $\mathrm{m}$ divides $\mathrm{b}$. By a similar argument we can see that $\mathrm{m}$ divides $\mathrm{a}$ as well. We have now reached a contradiction since $\mathrm{m}>1$ divides $\mathrm{a}$, $\mathrm{b}$, and $\mathrm{c}$ which are relatively prime. Therefore, the subtorus action must be free.

The free subtorus action rotates the last $k$ circles in the fibers of $\tilde{\mathcal{M}}^{n+k+3}$ which are parameterized by $(\phi^{n+1},\ldots,\phi^{n+k})$, while keeping the first $n$ circles fixed. Hence, viewing the spacetime as a bundle with torus fibers, the projection map $\tilde{\mathcal{M}}^{n+k+3} \to \tilde{\mathcal{M}}^{n+k+3}/T^k$ may be described by 
\begin{equation*}
(\mathbf{p}, \phi^1, \dots, \phi^{n+k}) \mapsto (\mathbf{p}, \phi^1, \dots, \phi^n),
\end{equation*}
where $\mathbf{p}\in\mathbb{R}_{+}^2$. To show that the quotient space is indeed homeomorphic to the given topology $\mathcal{M}^{n+3}$, we observe that the projection map implies that the rod data set $\tilde{\mathcal{D}}$ encoding the higher dimensional topology, descends down to the rod data set $\mathcal{D}$ for $\tilde{\mathcal{M}}^{n+k+3}/T^k$; this will be shown in detail below.

Lastly, since the free subtorus action is by isometries, and the static vacuum total space $\tilde{\mathcal{M}}^{n+k+3}$ is regular, the same is true of the quotient $\mathcal{M}^{n+3}$. In particular, this solution is devoid of conical singularities. We note that as a consequence of the dimensional reduction on tori, Kaluza-Klein matter fields will be present. Indeed, the metric on $\tilde{\mathcal{M}}^{n+k+3}$ may be expressed in Kaluza-Klein format as
\begin{equation*}
\tilde{\mathbf{g}}=\mathbf{g}+\sum_{\mu,\nu=n+1}^{n+k}h_{\mu\nu}(d\phi^{\mu}+A^{\mu}_i d\phi^i)(d\phi^\nu +A^{\nu}_j d\phi^j),
\end{equation*}
where $i,j=1,\ldots,n$, $h_{\mu\nu}=\tilde{f}_{\mu\nu}$, $h_{\mu\nu}A_i^{\mu}=\tilde{f}_{\nu i}$, and the (quotient) metric $\mathbf{g}$ on $\mathcal{M}^{n+3}$ is given in Weyl-Papapetrou \cite{Harmark} form by
\begin{equation*}
\mathbf{g}=-(fh)^{-1}\!\rho^2 dt^2 +(fh)^{-1}e^{2\sigma}(d\rho^2 \!+dz^2)+\sum_{i,j=1}^{n}f_{ij}d\phi^i d\phi^j,
\end{equation*}
with $f_{ij}+h_{\mu\nu}A_i^{\mu} A_j^{\nu}=\tilde{f}_{ij}$, $f=\det (f_{ij})$, and $h=\det(h_{\mu\nu})$.
The dimensionally reduced Lagrangian on $\mathcal{M}^{n+3}$ may then be expressed \cite[Section 11.4]{ortin} as
\begin{equation*}
\mathcal{L} = \sqrt{hg} \left(R - \frac{1}{4}(|\mathrm{Tr}(h^{-1}\nabla h)|^2 + \mathrm{Tr}(h^{-1}\nabla h)^2+|\mathcal{F}|^2)\right),
\end{equation*}
where $R$ is the scalar curvature of $\mathbf{g}$, $|\mathcal{F}|^2=h_{\mu\nu}\mathcal{F}^{\mu \mathbf{ij}}\mathcal{F}_{\mathbf{ij}}^{\nu}$ with $\mathcal{F}^{\mu}=dA^{\mu}$ and $\mathbf{i,j}$ labelling the coordinates of $\mathbf{g}$, and $g=-\det \mathbf{g}$. The second and third terms in the Lagrangian
give rise to the action for a sigma model (harmonic map) with target space $SL(k,\mathbb{R})/SO(k)$ (see \cite{KKRW}), while the fourth term yields the action for Abelian gauge fields. In partciular, the associated stress-energy-momentum tensor will satisfy the dominant energy condition. This property is verified, and the relevant equations of motion are given, in the appendix.  

To see directly that $\mathcal{D}$ is the rod data set for $\tilde{\mathcal{M}}^{n+k+3}/T^k$, consider an axis rod $\Gamma_l$ with rod structure $\tilde{\mathbf{v}}_l=\bar{\mathbf{v}}_l +\mathbf{e}_{n+l}$ within $\tilde{\mathcal{M}}^{n+k+3}$. Then $\tilde{f}_{mj}v_{l}^{j}+\tilde{f}_{m(n+l)}=0$ on the axis rod for $m=1,\ldots,n+k$. It follows from relations above that
\begin{equation*}
f_{ij}v_{l}^{j}+h_{\mu\nu}A_{i}^{\mu}A_{j}^{\nu}v_{l}^{j}=\tilde{f}_{ij}v_{l}^{j}=-\tilde{f}_{i(n+l)},
\end{equation*}
and $h_{\mu\nu}A_{j}^{\nu}v_{l}^{j}=\tilde{f}_{\mu j}v_{l}^{j}=-\tilde{f}_{\mu(n+l)}$. Therefore
\begin{equation*}
f_{ij}v_{l}^{j}=\tilde{f}_{\mu(n+l)}A_{i}^{\mu}-\tilde{f}_{i(n+l)}= h_{\mu(n+l)}A_{i}^{\mu}-\tilde{f}_{i(n+l)} =0,  
\end{equation*}
showing that $\mathbf{v}$ is the rod structure for $\Gamma_l$ within the quotient.

We now record what has been established. A globally hyperbolic spacetime of dimension $n+3$ will be referred to as \textit{multi-axisymmetric}, if a (noncompact) Cauchy slice admits the symmetry group $U(1)^n$ with a simply connected 2-dimensional orbit space, so that its topology is completely determined by an admissible rod data set $\mathcal{D}$.

\begin{theorem}
Any possible topology of the domain of outer communication for a multi-axisymmetric spacetime of dimension greater than or equal to 4, is realizable by a regular static solution of the Einstein equations with Kaluza-Klein matter. In particular, these solutions are obtained from a higher dimensional asymptotically Kaluza-Klein vacuum solution by dimensional reduction on tori.
\end{theorem}


The 5-dimensional case is of particular interest. By choosing rod structures $\v_1=(1,0)$ and $\v_k=(0,1)$ for the two semi-infinite rods $\Gamma_1$ and $\Gamma_k$, cross-sections of the time slice $M^4$ near spatial infinity will be 3-spheres, and in this region the spacetime curvature will approach zero; solutions with these two properties will be referred to as \textit{formally asymptotically flat}. It should be noted that various, often more specialized, notions of asymptotic flatness appear in the literature, which may not be applicable to the solutions discussed here. In particular, the 2-dimensional tori that foliate the $S^3$ cross-sections, which arise from the Hopf fibration, may not grow in the asymptotic end. We state the next result with a focus on the topology of black holes.

\begin{corollary}
There exist $5$-dimensional regular formally asymptotically flat static bi-axially symmetric solutions of the Einstein equations with Kaluza-Klein matter, supporting any finite configuration of nondegenerate black hole horizons of the form $S^3$, $S^1\times S^2$, or $L(p,q)$ where $1\leq q< p$ with $\mathrm{gcd}(p,q)=1$. 
\end{corollary}

We remark that all solutions discussed may be written down explicitly in terms of the harmonic functions $u_i$ and rod structures $\v_i$. Moreover, it is possible to replace static solutions with stationary solutions in these results, thus giving rotation to the constructed black holes.
This is accomplished by utilizing the harmonic map approach of \cite{KKRW}, instead of the harmonic function technique to obtain the relevant higher dimensional vacuum solutions. Although these stationary solutions may be shown to exist with the same underlying topologies of the static solutions described above, the angular momenta of the individual black holes cannot be fully prescribed due to the process of balancing conical singularities.

{\bf Appendix.}
Here it is shown that the Kaluza-Klein matter fields satisfy the dominant energy condition, and their equations of motion are also derived. Let $\mathcal{L}_M$ denote that matter portion of the Lagrangian $\mathcal{L}$, and consider the
stress-energy-momentum tensor defined by
\begin{align}\label{aofoaihogq}
\begin{split}
4\mathbf{T}_{\mathbf{ij}}=&-\frac{4}{\sqrt{hg}}\frac{\delta\mathcal{L}_M}{\delta \mathbf{g}^{\mathbf{ij}}}\\
=&\mathrm{Tr}(h^{-1}\partial_{\mathbf{i}}h)\mathrm{Tr}(h^{-1}\partial_{\mathbf{j}}h)
+\mathrm{Tr}(h^{-1}\partial_{\mathbf{i}}h h^{-1}\partial_{\mathbf{j}}h)\\
&-\frac{1}{2}\left(|\mathrm{Tr}(h^{-1}\nabla h)|^2 + \mathrm{Tr}(h^{-1}\nabla h)^2\right)\mathbf{g}_{\mathbf{ij}}\\
&+2h_{\mu\nu}g^{\mathbf{lm}}\mathcal{F}_{\mathbf{il}}^{\mu}\mathcal{F}_{\mathbf{jm}}^{\nu}
-\frac{1}{2}|\mathcal{F}|^2 \mathbf{g}_{\mathbf{ij}}.
\end{split}
\end{align}
To confirm that $\mathbf{T}(X,Y)\geq 0$ for all future-pointing $\mathbf{g}$-causal vectors $X$ and $Y$, it suffices to establish that 
\begin{equation}\label{aeofoiqahoihq}
\mathbf{T}(\mathbf{n},\mathbf{n})\geq |\mathbf{T}(\mathbf{n},\cdot)|
\end{equation}
for any unit timelike vector $\mathbf{n}$, where $\cdot$ represents (spacelike) vectors orthogonal to $\mathbf{n}$ having norm less than or equal to 1. Observe that the last line of \eqref{aofoaihogq} may be interpreted as a sum of stress-energy-momentum tensors for abliean gauge fields, each of which satisfies the dominant energy condition (see 
\cite{AKK}). Therefore it is enough to show \eqref{aeofoiqahoihq} for the remaining piece $\mathbf{T}_H$ of $\mathbf{T}$. To this end, we compute
\begin{align*}
\begin{split}
\mathbf{T}_{H}(\mathbf{n},\mathbf{n})=&\frac{1}{2}\left(|\mathrm{Tr}(h^{-1}\partial_{\mathbf{n}}h)|^2
+\mathrm{Tr}(h^{-1} \partial_{\mathbf{n}}h)^2\right)\\
&+\frac{1}{2}\left(|\mathrm{Tr}(h^{-1}\vec{\nabla}h)|^2
+\mathrm{Tr}(h^{-1}\vec{\nabla} h)^2\right)
\end{split}
\end{align*}
where $\vec{\nabla}$ denotes derivatives in directions perpendicular to $\mathbf{n}$,
and
\begin{equation*}
\mathbf{T}_{H}(\mathbf{n},\cdot)=\mathrm{Tr}(h^{-1}\partial_{\mathbf{n}}h)\mathrm{Tr}(h^{-1}\vec{\nabla}h)
+\mathrm{Tr}(h^{-1}\partial_{\mathbf{n}}h h^{-1}\vec{\nabla}h).
\end{equation*}
As mentioned previously, the second and third terms of $\mathcal{L}$ constitute the action for a sigma model
with target Riemannian symmetric space $SL(k,\mathbb{R})/SO(k)$. Therefore, the expressions for $\mathbf{T}_H(\mathbf{n},\mathbf{n})$ and $\mathbf{T}_{H}(\mathbf{n},\cdot)$ consist of inner products between tangent vectors to the target space. It follows that the Cauchy-Schwarz inequality may be applied to obtain the desired inequality \eqref{aeofoiqahoihq} for $\mathbf{T}_H$. 

Lastly, to find the equations of motion for the matter fields, it is convenient to conformally change the spacetime metric on $\mathcal{M}^{n+3}$, namely set $\mathbf{g}=e^{2\psi}\bar{\mathbf{g}}$ where $\psi=-\tfrac{1}{2(n+1)}\log\det h$. The Lagrangian then becomes
\begin{equation*}
\mathcal{L}\! =\! \sqrt{\bar{g}}\! \left(\!\!\bar{R}\! -\! \frac{1}{4}(c_n|\mathrm{Tr}(h^{-1}\nabla h)|^2 \!+\! \mathrm{Tr}(h^{-1}\nabla h)^2 \!+ \! e^{-2\psi}|\mathcal{F}|^2)\!\right)\!,
\end{equation*}
where $\bar{g}=\det\bar{\mathbf{g}}$, $\bar{R}$ is the scalar curvature of $\bar{\mathbf{g}}$, $c_n=\tfrac{2n+3}{n+1}$, and all norms are now with respect to $\bar{\mathbf{g}}$. A straightforward, albeit tedious computation, then shows that the associated Euler-Lagrange equations are 
\begin{equation*}
\bar{\Box} h_{\mu\nu}-h^{\iota\kappa}\bar{\nabla} h_{\mu\iota} \cdot\bar{\nabla}h_{\kappa\nu} =\bar{c}_n e^{-2\psi}|\mathcal{F}|^2 h_{\mu\nu},
\end{equation*}
\begin{equation*}
\bar{\mathrm{div}}\left(e^{-2\psi} h_{\mu\nu}\mathcal{F}^{\nu}\right)=0,
\end{equation*}
where $\bar{\Box}$ and $\bar{\mathrm{div}}$ are the wave and divergence operators with respect to $\bar{\mathbf{g}}$, and $\bar{c}_n^{-1}=4(n^2+5n+5)$.

{\bf Acknowledgments.} M. Khuri acknowledges the support of NSF Grant DMS-2104229. 
The authors would like to thank Hari Kunduri, Martin Rocek, and Phil Saad for helpful comments.

\end{document}